\documentstyle[12pt]{article}
\makeatletter
\@addtoreset{equation}{section}

\makeatother

\begin{document}

\begin{flushright}
Jun 2002

KEK-TH-830
\end{flushright}

\begin{center}

\vspace{5cm}

{\Large Deformation of CHS model}

\vspace{2cm}

Takao Suyama \footnote{e-mail address : tsuyama@post.kek.jp}

\vspace{1cm}

{\it Theory Group, KEK}

{\it Tsukuba, Ibaraki 305-0801, Japan}

\vspace{4cm}

{\bf Abstract} 

\end{center}

We calculate mass spectrum of CHS model deformed by an exactly marginal operator, and find that 
there are tachyons which are not localized in the target space. 
Similar deformation is discussed in another CFT which corresponds to separated NS5-branes. 
A condensation of the tachyons is briefly argued. 

\newpage

\vspace{1cm}

\section{Introduction}

\vspace{5mm}

Perturbative vacua of string theory are often unstable if there is no spacetime supersymmetry. 
In many cases, one would find tachyonic states in the mass spectrum, which signal the instability of 
the vacua. 
Then it would be natural to ask what happens when the tachyons acquire nonzero vevs. 
To analyze this kind of process, one would have to consider an off-shell dynamics of string theory 
since the turning on such vevs would not be an on-shell deformation of the background. 
In the case of open string tachyon condensation, this is achieved by using string field theory 
\cite{SFT}\cite{SFT2}\cite{SFT3}\cite{SFT4}. 
There are also some attempts to discuss condensation of closed string tachyons by using 
D-brane probes \cite{APS} and 
renormalization group of two-dimensional theories 
\cite{Vafa}\cite{HKMM}\cite{TP}\cite{DGHM}. 
The research on the closed string tachyon condensation may provide us a key to understand a proper 
formulation of (closed) string theory beyond perturbative definition. 

In this couple of years, two types of perturbative backgrounds which produce closed string tachyons 
are studied intensively. 

The one is the geometry whose metric is 
\begin{equation}
ds^2 = \eta_{\mu\nu}dx^\mu dx^\nu + r^2(d\theta+qdy)^2+dy^2. 
    \label{Melvin}
\end{equation}
The dimensional reduction along the $y$-direction, which is compactified on $S^1$, gives a curved 
background with a magnetic flux, which is known as Melvin background \cite{Melvin}. 
String theories on this background (\ref{Melvin}) are solvable \cite{RT}, and 
it was conjectured that superstring theories on this background are equivalent to ten-dimensional 
non-supersymmetric string theories in appropriate limits \cite{TypeII}\cite{hetero}. 
The condensation of the tachyons was discussed \cite{TypeII}\cite{DGHM}\cite{TC}, 
and it was conjectured that the endpoint of the 
condensation corresponds to the flat supersymmetric background. 

The other tachyonic backgrounds are non-supersymmetric orbifolds \cite{APS}. 
In this case, closed string tachyons come from twisted sectors, and thus, the tachyons are localized 
at the orbifold singularity. 
It was conjectured that the condensation of the tachyons resolves the singularity. 
The effect of the condensation would not confined around the singularity, and finally, the orbifold 
would be deformed to become the flat supersymmetric background. 

Note that the tachyons in the Melvin case are also localized at $r=0$, and thus, the recent 
investigations are focused on the tachyons which are localized in target spacetime. 

\vspace{2mm}

In this paper, we will consider a CFT which is known as CHS model \cite{CHS}. 
This CFT describes Type II strings in the vicinity of NS5-branes. 
We will show that there appear tachyons in the CHS model when it is deformed by a marginal operator 
which breaks the spacetime supersymmetry completely. 
The tachyons would not be localized in spacetime. 
Therefore, condensation of these tachyons might be quite different from the other tachyon condensation 
studied before. 
We will briefly discuss a condensation of the tachyons, in analogy with the 'tachyon condensation' in 
the Liouville theory. 

This paper is organized as follows. 
In section \ref{review}, we briefly review the CHS model. 
A deformation of the CHS model is investigated in section \ref{deformedCHS}. 
In this section, the mass spectrum of the deformed theory is calculated exactly, and it is shown that 
there are tachyonic states. 
There is another CFT related to NS5-branes which are separated from each other \cite{OV}. 
The corresponding deformation in this related model is also discussed in section \ref{OVcase}. 
In section \ref{proposal}, we propose a mechanism to stabilize the tachyonic instability.

\vspace{1cm}

\section{CHS model}  \label{review}

\vspace{5mm}

In this section, we will briefly review an exactly solvable CFT which describes string theory 
in the vicinity of NS5-branes \cite{CHS}. 

\vspace{2mm}

The supergravity solution for coinciding NS5-branes in Type II theory is given as follows 
\begin{eqnarray}
ds^2 &=& \eta_{\mu\nu}dx^\mu dx^\nu +e^{2\Phi}\delta_{mn}dx^mdx^n, \nonumber \\
H_{lmn} &=& {\epsilon^j}_{lmn}\partial_j\Phi, 
     \label{SUGRA} \\
e^{2\Phi} &=& e^{2\Phi_0}+\frac Q{r^2}. \nonumber 
\end{eqnarray}
Here $x^\mu$ are the tangential directions to the NS5-branes, and $x^m$ are the transverse directions. 
The parameter $Q$ is proportional to the number $N$ of the NS5-branes. 
This solution preserves sixteen supercharges. 

The dynamics of strings in the NS5-brane background can, in principle, be studied by considering a 
non-linear sigma model whose background metric and fields are given by (\ref{SUGRA}). 
However, the resulting sigma model is too complicated to analyze. 
The model becomes very simple when we focus on a region where the constant $e^{2\Phi_0}$ is 
negligible, i.e. where $r$ is small. 
In this region, the metric and the dilaton become 
\begin{eqnarray}
ds^2 &=& \eta_{\mu\nu}dx^\mu dx^\nu+d\rho^2+Qd\Omega_3^2, \nonumber \\
\Phi &=& -\frac1{\sqrt{Q}}\rho+\frac12\log Q, 
    \label{throat}
\end{eqnarray}
where $\rho=\sqrt{Q}\log r$. 
Therefore, the background becomes ${\bf R}^{1,5}\times {\bf R}\times S^3$ with the dilaton which 
depends on the coordinate $\rho$ of ${\bf R}$ linearly. 

\vspace{2mm}

The sigma model on this background corresponds to a product of solvable CFT's. 
${\bf R}^{1,5}$ corresponds to the free CFT, ${\bf R}$ to the linear dilaton CFT or Feigin-Fuchs 
system and $S^3$ to the SU(2) WZNW model. 
Details of the latter two CFT's are specified by $N$; the central charge $c$ of the linear dilaton 
CFT is $c=1+6/N$, and the level $k$ of the SU(2) current algebra is $k=N-2$ \cite{CHS}. 
Note that the total central charge of this system is 
\begin{equation}
6+\left(1+\frac6N\right)+\frac{3(N-2)}{(N-2)+2}+5 = 15,
\end{equation}
(5 comes from free fermions) and thus this background is appropriate for Type II superstring. 

In this background, the dilaton diverges as $\rho \to -\infty$. 
This means that the above CFT description would not be available in this region. 
There exists a range of $\rho$ in which the throat approximation (\ref{throat}) is valid and the 
string coupling is 
small, when the asymptotic value $e^{\Phi_0}$ of the string coupling is small enough.

\vspace{1cm}

\section{A marginal deformation}   \label{deformedCHS}

\vspace{5mm}

The CHS model can be deformed by an exactly marginal operator 
\begin{equation}
\delta S \propto \int d^2z (J^3+\psi\psi^\dag)(\tilde{J}^3+\tilde{\psi}\tilde{\psi}^\dag), 
    \label{deform}
\end{equation}
where $J^a\ (a=1,2,3)$ are the SU(2) currents 
\begin{equation}
J^a(z)J^b(w) \sim \frac{k/2}{(z-w)^2}\delta^{ab}+\frac{i\epsilon^{abc}}{z-w}J^c(w), 
\end{equation}
and $\psi$ is a complex fermion which is constructed 
from two superpartner fermions of the SU(2) currents. 
The currents and fields with tilde are in the right-moving sector. 
This kind of deformation was discussed in the bosonic WZNW model \cite{JJdeform}, and in 
superstrings \cite{deformhetero}. 
Although the CHS model has N=4 worldsheet supersymmetry, the deformation (\ref{deform}) preserves 
only N=1 supersymmetry. 
Therefore the spacetime supersymmetry is completely broken by the deformation. 

This deformation is interesting since, as we will show below, it is possible to analyze the deformed 
system with any finite value of the deformation parameter. 

\vspace{5mm}

\subsection{Sigma model}

\vspace{3mm}

Since N=1 supersymmetry is preserved, one can interpret the deformation (\ref{deform}) as a change of 
the background metric and fields. 
In this subsection, we consider the sigma model metric at tree-level. 
Roughly speaking, the deformation distorts the $S^3$ geometry. 
One can find the resulting metric on the $S^3$ by using non-linear sigma model of $S^3$. 
Such investigation was done in \cite{JJdeform}, and the deformed metric is 
\begin{eqnarray}
ds_{\alpha}^2 &=& k\left[ d\phi^2
  +\frac1\Delta\left\{(\cos\alpha-k\sin\alpha)^2\sin^2\frac\phi 2+\cos^2\alpha\cos^2\frac\phi 2\right\}
   (d\theta_L^2+d\theta_R^2) \right. \nonumber \\
& & \left.
    +\frac2\Delta\left\{-(\cos\alpha-k\sin\alpha)^2\sin^2\frac\phi 2
    +\cos^2\alpha\cos^2\frac\phi 2\right\}d\theta_Ld\theta_R\right], 
              \label{Dmetric}
\end{eqnarray}
where $\phi,\theta_{L(R)}$ are angular variables parametrizing SU(2) elements as 
\begin{equation}
g=e^{i\theta_L\sigma^2/2}e^{i\phi\sigma^1/2}e^{i\theta_R\sigma^2/2}. 
\end{equation}
In (\ref{Dmetric}), $\alpha$ is the deformation parameter, and 
\begin{equation}
\Delta = \cos^2\alpha\sin^2\frac\phi 2+(\cos\alpha-k\sin\alpha)^2\cos^2\frac\phi 2. 
\end{equation}
The angular variables take the values in 
\begin{equation}
0\leq \theta_{L(R)} <2\pi, \hspace{1cm} 0\leq \phi \le \pi.
\end{equation}
The undeformed metric corresponds to $\alpha=0$, and thus
\begin{equation}
ds_{\alpha=0}^2 = k\left[ d\phi^2+\sin^2\phi d\theta_L^2+(d\theta_R+\cos\phi d\theta_L)^2\right]. 
\end{equation}
This metric indicates $S^3$ as an $S^1$ fibration over $S^2$. 
In the $\alpha\ne 0$ case, the base $S^2$ is squeezed, and the radius of the fiber $S^1$ 
depends on $\phi$.

\vspace{5mm}

\subsection{Mass spectrum of the deformed model}

\vspace{3mm}

It is very remarkable that the mass spectrum of the deformed model can be calculated exactly. 
This is due to the fact that the currents in the deformation term (\ref{deform}) can be rewritten by 
using free bosons 
\begin{equation}
J^3 = i\sqrt{\frac {N-2}2}\partial H, \hspace{1cm} \psi\psi^\dag = i\partial \varphi, 
\end{equation}
where the free bosons are normalized as 
\begin{equation}
H(z)H(w)\sim -\log(z-w), \hspace{1cm} \varphi(z)\varphi(w)\sim -\log(z-w). 
\end{equation}
The other SU(2) currents are constructed by using $H$ and the parafermion. 
Thus the deformation (\ref{deform}) is just a deformation which changes the radius of the free boson 
$Y\propto\sqrt{(N-2)/2}H+\varphi$. 

Recall that such a deformation induces an SO(1,1) transformation of the momentum 
\begin{equation}
\left(
\begin{array}{c}
 p_L' \\ p_R' 
\end{array}
\right) = \left(
\begin{array}{cc}
 \cosh x & \sinh x \\ \sinh x & \cosh x 
\end{array}
\right)\left(
\begin{array}{c}
 p_L \\ p_R 
\end{array}
\right).
\end{equation}
Here $x$ is the deformation parameter. 
In our case, $p_L=j^3+n_F,\ p_R=\tilde{j}^3+\tilde{n}_F$ and $j^3,n_F$ are eigenvalues of 
$J^3,\psi\psi^\dag$, respectively. 
Similar calculation was performed for bosonic WZNW model in \cite{bosonicPF}. 

Now we obtain the exact formula for mass shift 
\begin{eqnarray}
\frac{\alpha'}4\delta M^2 
&=& \frac1N(p_L'^2+p_R'^2)-\frac1N(p_L^2+p_R^2) \nonumber \\ 
&=& \frac1N\left[\left\{(j^3+n_F)^2+(\tilde{j}^3+\tilde{n}_F)^2\right\}\sinh^2x
   +(j^3+n_F)(\tilde{j}^3+\tilde{n}_F)\sinh2x\right]. \nonumber \\
          \label{shift}
\end{eqnarray}
This quantity can be negative since the $\sinh 2x$ term can be negative. 
The lowest possible value of the mass shift is 
\begin{equation}
\frac{\alpha'}4\delta M^2 \ge -\frac1N \min \left[ (j^3+n_F)^2, (\tilde{j}^3+\tilde{n}_F)^2 \right]. 
\end{equation}

Let us consider states of the form 
\begin{equation}
|free\rangle \otimes |\beta\rangle\otimes |j,j^3\rangle, 
        \label{states}
\end{equation}
where $|\beta\rangle$, corresponding to the vertex operator $e^{\beta\phi}$, and $|j,j^3\rangle$ 
are primary states of the linear dilaton CFT and SU(2) WZNW, respectively. 
Here $j$ is the spin of the primary states of the SU(2) WZNW. 
The masses of such states in the NS-NS sector can be calculated exactly 
\begin{equation}
\frac{\alpha'}4 M^2 = -\frac12+\frac1{4N}+\frac{j(j+1)}N+\frac12n_F^2
                            +\frac{\alpha'}4\delta M^2 + \mbox{(osc.)}
\end{equation}
The last term indicates contributions which come from oscillators of free fields. 

For example, consider a state with 
\begin{equation}
j=j^3=\tilde{j}=\tilde{j}^3=0,\ n_F=\tilde{n}_F=1, 
     \label{tachyon}
\end{equation}
and no oscillation mode is excited
\footnote{The $n_F$=$\tilde{n}_F$=$0$ case is excluded by the GSO projection. }. 
Then the mass of this state is given as follows 
\begin{equation}
\frac{\alpha'}4M^2 = \frac1N\left(-\frac34+e^{2x}\right). 
\end{equation}
Therefore, this state becomes tachyonic if $e^{2x}<3/4$. 
Note that the spectrum is invariant under $x\to -x$, which is just the T-duality. 
Thus tachyons also appear for large $x$. 

The state (\ref{tachyon}) is not the most tachyonic state of the system. 
To find them, it would be helpful to see a part of the energy-momentum tensor 
\begin{eqnarray}
T &=& -\frac12(\partial H)^2-\frac12(\partial \varphi)^2 +\cdots \nonumber \\
  &=& -\frac1N\left(\sqrt{\frac{N-2}2}\partial H+\partial \varphi\right)^2
      -\frac1{2N(N-2)}\left(2\sqrt{\frac{N-2}2}\partial H-(N-2)\partial\varphi\right)^2+\cdots. 
           \nonumber \\
\end{eqnarray}
The mass shift (\ref{shift}) is independent of the zero mode of $\sqrt{2(N-2)}H-(N-2)\varphi$. 
This implies that the most tachyonic states would satisfy 
\begin{equation}
2j^3-(N-2)n_F=0. 
    \label{neutral}
\end{equation}
For such a state with $n_F=\tilde{n}_F=1$, the mass is 
\begin{equation}
\frac{\alpha'}4 M^2 = -\frac12+\frac1{4N} \le -\frac3{4N} 
\end{equation}
for $x\to -\infty$. 
Therefore, this state is more tachyonic than the state (\ref{tachyon}). 

One can show that no spacetime fermion corresponding to the state (\ref{states}) discussed above 
becomes tachyonic by the deformation. 

\vspace{2mm}

The full spectrum can be easily analyzed for $N=3$, i.e. the level 1 case 
\begin{eqnarray}
\frac{\alpha'}4 M^2 
&=& E_0+\frac1{4\cdot3}+(n+j)^2+\frac12n_F^2+\mbox{(osc.)}+\frac{\alpha'}4\delta M^2
                      \nonumber \\
&=& E_0+\frac1{12}+\frac16(2n+2j-n_F)^2+\mbox{(osc.)} \nonumber \\
& &   +\frac13\left[(n+j+n_F)\sinh x+(\tilde{n}+j+\tilde{n}_F)\cosh x\right]^2. 
\end{eqnarray}
In this case, $j=0,1/2$ and $n,\tilde{n}\in {\bf Z}$. 
In the NS-NS sector, $E_0=-1/2$ and the most tachyonic states are the states with $2j^3=2n+2j=n_F$, 
which has solutions since $n_F\in {\bf Z}$. 
Then the smallest mass squared is 
\begin{equation}
\frac{\alpha'}4M^2 = -\frac5{12}. 
\end{equation}
On the other hand, in the R-NS sector, $E_0=-\frac18$ and $n_F\in {\bf Z}+1/2$, and thus the lightest 
states are massless states. 
Similar calculation shows that there is no tachyonic state in the R-R sector. 

\vspace{2mm}

It should be noted that there is no localized state in the deformed CHS model. 
Before the deformation, the target spacetime is ${\bf R}^{1,5}\times {\bf R}\times S^3$. 
Therefore, there is no localized strings. 
And, as shown above, the deformation (\ref{deform}) does not change the number of zero modes. 
Thus originally delocalized states are still delocalized, and the tachyons are not localized in the 
target spacetime.

\vspace{1cm}

\section{CFT description without strong coupling region}   \label{OVcase}

\vspace{5mm}

It is known that there is a weakly coupled CFT description of NS5-branes \cite{OV}. 
The NS5-branes in this case are placed on a circle symmetrically. 
The corresponding CFT is an orbifold of a product of solvable CFT's 
\begin{equation}
{\bf R}^{1,5}\times \left[\ \mbox{SL(2)/U(1)}\times\mbox{SU(2)/U(1)}\ \right]/G. 
       \label{coset}
\end{equation}
Here the first part represents the free CFT, the second corresponds to a CFT of the two-dimensional 
Euclidean black hole \cite{2dimBH} and the last one indicates an N=2 minimal model. 
The orbifolding by $G$ is necessary to impose the GSO projection. 
The string coupling is bounded from above by a constant $g_0$. 
This $g_0$ is related to the radius $r$ of the circle on which the NS5-branes are placed. 
The constant $g_0$ becomes smaller when $r$ becomes larger. 
Thus the above CFT in the $r\to 0$ limit will correspond to the CHS model. 

There is another description of the separated NS5-branes \cite{LST}. 
It was shown in \cite{duality} that supersymmetric coset SL(2)/U(1) is equivalent to N=2 Liouville 
theory, 
which is a two-dimensional theory with the target space ${\bf R}\times S^1$ and with a potential 
\begin{equation}
\lambda\int d^2\theta e^{-\frac1Q(\phi+iY)}+\mbox{h.c.}
   \label{potential}
\end{equation}
Here $\lambda$ is a coupling constant. 
The CHS model limit corresponds to the limit $\lambda\to 0$. 
This can be easily shown as follows. 

Recall the fields in the CHS model. 
There are the SU(2) currents $J^a$ with level $k$, a non-compact boson $\phi$ with a background 
charge and four free fermions. 
Take $J^a$ and two free fermions, and construct the following currents \cite{SCA}, 
\begin{eqnarray}
T &=& T^{SU(2)}+T^{\psi}-\frac1{k+2}(J^3+\psi\psi^\dag)^2, \nonumber \\
G^+ &=& \sqrt{\frac2{k+2}}\psi^\dag J^+, \nonumber \\
G^- &=& \sqrt{\frac2{k+2}}\psi J^-, 
      \label{algebra}     \\
J &=& \frac1{k+2}(2J^3-k\psi\psi^\dag), \nonumber \\
T^{U(1)} &=& \frac1{k+2}\left(J^3+\psi\psi^\dag\right)^2. \nonumber 
\end{eqnarray}
Here $T^{SU(2)},T^{\psi}$ are ordinary energy-momentum tensors of the SU(2) current algebra and 
the complex fermion $\psi$, respectively. 
One can easily show that $T,G^{\pm},J$ are the generators of the N=2 superconformal algebra, and 
$T^{U(1)}$, which is the energy-momentum tensor of a compact boson, is decoupled from them. 
The central charge of the N=2 system is $3k/(k+2)$, and thus this is the N=2 minimal model. 
Therefore, the CHS model can be rewritten as a product of ${\bf R}\times S^1$ CFT with a background 
charge and the N=2 minimal model, which is equivalent to the product of the N=2 Liouville theory 
and the N=2 minimal model, with vanishing potential. 

In the Liouville description, the potential (\ref{potential}) prevents strings from going into the 
strong coupling region. 
On the other hand, in the coset description, the strong coupling region is cut out, and the target 
space has the cigar geometry. 

\vspace{2mm}

By the above reconstruction of the CHS model, it becomes manifest that the deformation (\ref{deform}) 
corresponds to a change of the radius of a circle, and is exactly marginal. 
In the presence of the potential (\ref{potential}), however, the situation changes. 
That is, the radius-changing deformation should be accompanied by a modification of the form of the 
potential, in order for the periodicity of the compact boson to be consistent with the periodicity of 
the potential. 
Then the weight of the potential would change and the system would leave the conformal fixed 
point\footnote{I would like to thank T.Eguchi for clarifying this point.}. 
Therefore, in the separated NS5-brane case, the corresponding deformation would not be marginal.

\vspace{1cm}

\section{Discussion}    \label{proposal}

\vspace{5mm}

We have shown the appearance of tachyonic states in the CHS model deformed by a marginal operator. 
This kind of deformation is not marginal when one consider a different NS5-brane configuration whose 
CFT description is weakly coupled. 
It is due to the presence of the Liouville-like potential (\ref{potential}) which effectively cuts out 
the strong coupling region and, at the same time, makes the deformation be off-shell. 

It is interesting to remember that, in the Liouville theory, the potential can be regarded as the one 
generated by a 'tachyon condensation'. 
Thus, naively, one would expect that a similar potential is generated in the deformed CHS model by a 
condensation of the tachyonic states discussed in section \ref{deformedCHS}. 

Let us recall the situation in the Liouville theory. 
The linear dilaton CFT is solvable, but the string coupling becomes arbitrary large. 
Therefore it would not be suitable for string perturbation theory. 
One can instead consider the linear dilaton background with a tachyon profile, by adding the tachyon 
vertex operator to the worldsheet Lagrangian. 
Then the vertex operator acts as a potential which prevents strings from going into the strong coupling 
region. 
This situation would be very similar to the CHS model. 

One interesting observation is that the most tachyonic states are neutral under the U(1) current of 
the N=2 minimal model. 
As explained in section \ref{deformedCHS}, the most tachyonic states would satisfy the condition 
(\ref{neutral}). 
And the quantity $2j^3-(N-2)n_F$ is proportional to the charge of the U(1) current $J$ in 
(\ref{algebra}). 
Therefore, the relevant part of the vertex operators of the most tachyonic states would be the 
${\bf R}\times S^1$ part, and thus, it would be natural to expect that a condensation of such states 
generates the Liouville-like potential. 

Let us consider the deformed CHS model with a 'tachyon background', i.e. 
${\bf R}\times S^1\times$SU(2)/U(1) model with the potential 
\begin{equation}
\lambda\int d^2\theta e^{-\left(\frac1Q\phi +i\frac1RY\right)} + \mbox{h.c.} 
   \label{potential2}
\end{equation}
with $R>Q$. 
In this case, the potential breaks the conformal symmetry. 
Therefore this background would be still unstable. 
To see what happens for this background, it would be convenient to discuss in its dual coset 
description (\ref{coset}). 

The radius $R$ of the $S^1$ of the N=2 Liouville theory is inversely proportional to the radius $R'$ of 
the asymptotic circle of the cigar geometry 
\begin{eqnarray}
ds^2 &=& k'\left[ d\rho^2+\tanh^2\rho d\theta^2\right], \nonumber \\
\Phi &=& \Phi_0-2\log\cosh \rho,  \nonumber 
\end{eqnarray}
in which $R'=\sqrt{k'}$. 
This is the geometry of the two-dimensional Euclidean black hole \cite{2dimBH}, and its temperature 
is inversely proportional to $R'$. 
Therefore, the above potential (\ref{potential2}) would correspond to the black hole with higher 
temperature than that of the stable black hole. 
Then it seems natural that the temperature of the system lowers by a radiation, and the system 
would be stabilized. 

This process could be described in a different way. 
By changing $R'$, there would appear a conical singularity at $r=0$. 
Then, twisted strings on the singularity would provide localized tachyonic states. 
In particular, if $R'=\sqrt{k'}/m$ for a positive integer $m$, the local geometry around the 
singularity would be just what is discussed in \cite{APS}. 
According to \cite{APS}, the condensation of the localized tachyons resolves the singularity, and it 
also deforms the whole geometry including the region far from the tip of the cigar. 
Note that the appearance of tachyons localized at the tip was discussed in a different context 
\cite{cigartachyon}. 

The above discussion thus seems to suggest that it would be possible to stabilize the tachyonic 
instability which appears in the deformed CHS model. 
The resulting system would be described eventually by the same CFT with the one which describes the 
separated NS5-branes. 
Therefore, if it is taken seriously, one may conclude that the tachyon condensation induces the 
separation of the NS5-branes. 
This may be related to the process discussed in \cite{HKMM}.

\vspace{1cm}

{\Large {\bf Acknowledgements}}

\vspace{5mm}

We would like to thank D.Bak, T.Eguchi, F.Lin, S.Minwalla, N.Ohta, J-H.Park, S.Sin, P.Yi, T.Yoneya 
for valuable discussions and comments. 
We would also like to thank Korea Institute for Advanced Study for kind hospitality.

\end{document}